\documentclass[prd,aps,preprint,amsmath,nofootinbib,amssymb,eqsecnum]{revtex4-1}
\usepackage{amsmath}
\usepackage{verbatim,graphics,graphicx,color,slashed}
\usepackage{hyperref}

\begin{document}

\title{\Large New Scalar Contributions to  $h\to Z\gamma$}
\author{Chian-Shu Chen$^{2}$\footnote{chianshu@phys.sinica.edu.tw}, Chao-Qiang Geng$^{1,2}$\footnote{geng@phys.nthu.edu.tw}, Da Huang$^{1}$\footnote{dahuang@phys.nthu.edu.tw}, and Lu-Hsing Tsai$^{1}$\footnote{lhtsai@phys.nthu.edu.tw}}
\affiliation{$^{1}$Department of Physics, National Tsing Hua University, Hsinchu, Taiwan
\\$^{2}$Physics Division, National Center for Theoretical Sciences, Hsinchu, Taiwan
}

\date{\today}

\begin{abstract}
We calculate the Higgs decay rate of  $h\to Z\gamma$ by including the contributions from new scalars with
arbitrary quantum numbers of the weak isospin ($T$) and hypercharge ($Y$)  in the standard model.
We find that our general formula for the decay rate of $h\to Z\gamma$ matches with that for
$h\to \gamma\gamma$ in the limit of $m_Z=0$, but it is different from those in the literature.
To illustrate our result, by taking the current $2\sigma$ excess of the $h\to \gamma\gamma$
rate measured by the LHC, we  examine the corresponding shift for the $Z\gamma$ decay channel
due to the new scalar. We show that the enhancement or reduction of the $h\to Z\gamma$ rate only
depends on the relative size of $T$ and the absolute value of $Y$.
Explicitly, we predict $0.76<R_{Z\gamma}\equiv \Gamma(h\to Z\gamma)/\Gamma_{SM}(h\to Z\gamma)<2.05$
by imposing the observed range of $1.5<R_{\gamma\gamma}\equiv \Gamma(h\to Z\gamma\gamma)/\Gamma_{SM}(h\to\gamma\gamma)<2$,
which is independent of the number of multiplets and the couplings to the Higgs particle as long as the scalars are heavier
than 200~GeV.  This result provides a clear signature for the future LHC measurements to test  physics beyond the standard model.

\end{abstract}

\maketitle

\section{Introduction}
\label{sec:intro}
The discovery of the Higgs-like particle ($h$) by ATLAS~\cite{atlas:2012gk} and CMS~\cite{cms:2012gu} collaborations is a great
triumph of particle physics. The following immediate question is whether this newly discovered particle is really the Higgs
particle in the Standard Model (SM). To achieve this goal, one has to determine its couplings to all the SM particles.
However, recent results from Higgs searches at the LHC have already shown a hint of  new physics in
the Higgs diphoton decay channel. The experimental data have shown that the diphoton decay rate is about $1.5 - 2.0$ times larger
than the SM prediction, while the measurements in other channels including $WW^*$ and $ZZ^*$ agree with the SM Higgs
 properties. Due to the fact that the Higgs particle is electric neutral, its coupling to diphoton must be induced by
some charged particles running in the loops, while in the SM, the W-boson and top-quark loops give the dominant
contributions~\cite{shifman,Ellis:1975ap,Ioffe:1976sd,Rizzo:1979mf,Resnick:1973vg}. If the enhancement in the diphoton channel persists,
it clearly indicates~\cite{calderone,Han:2003gf,Chen:2006cs,Cacciapaglia:2009ky,Cao:2011pg,Dawson:2012di,Kitahara:2012pb,Joglekar:2012vc,Hashimoto:2012qe,Delgado:2012sm,An:2012vp,ArkaniHamed:2012kq,Almeida:2012bq,Bertuzzo:2012bt,Moreau:2012da,Chala:2012af,Dawson:2012mk,Choi:2012he,Huo:2012tw} that there must be some additional new charged particles which couple to the Higgs and mediate
this decay process.
Based on the electroweak (EW) $SU(2)_{L}\times U(1)_{Y}$ gauge symmetry,
 these new particles should also contribute to the $h \rightarrow Z\gamma$ channel and generically lead to a shift in the
$Z\gamma$ decay width from the SM expectation.
 As pointed out in Refs.~\cite{Djouadi:1996yq,Djouadi:2005gi,Carena:2012xa,Chiang:2012qz,
Picek:2012ei,Huang:2012rh,Li:2012mu}, the simultaneous measurements of the $\gamma\gamma$ and $Z\gamma$ channels
at the LHC will
provide us with valuable information about the structure of new physics.

In this paper, we focus on the simple scenario in which extra contributions to the $\gamma\gamma$ and $Z\gamma$
decay widths arise from some new scalars beyond the SM particle content. Some similar scenarios
have already been explored in Refs.~\cite{Djouadi:1996yq,Djouadi:2005gi,Carena:2012xa,Chiang:2012qz,
Picek:2012ei,Huang:2012rh,Li:2012mu,Han:2012dd, Batell:2011pz,Baek:2012ub,Chang:2012ta,Melfo:2011nx,Arhrib:2011vc,Akeroyd:2012ms,Chun:2012jw,Dev:2013ff}. However, the formulae for
the scalar contributions to the $Z\gamma$ decay width used in the literature~\cite{Djouadi:1996yq,Djouadi:2005gi,Carena:2012xa,
Chiang:2012qz,Picek:2012ei,Huang:2012rh} are not consistent with each other.
Moreover, they cannot be reduced to the corresponding rates for $h\to\gamma\gamma$
 by taking the limit of $m_Z=0$ and  making the charge replacement.
As a result,  these numerical
predictions for the $Z\gamma$ decay width may not be reliable.
Clearly, it is timely important to re-calculate the scalar contributions to the decay.

This paper is organized as follows.
In section~\ref{sec:2}, we show the detail calculations of the scalar contributions to the decay rate of $h\to Z\gamma$.
In section~\ref{sec:3}, we give phenomenological analyses of the correlations between the $\gamma\gamma$ 
and $Z\gamma$ channels in several specific sets  as well as a generic one of the scalar multiplets. 
We summarize our results in section~\ref{sec:4}.

\section{Calculations of Scalar Contributions to $h\to Z\gamma$}
\label{sec:2}
For a general scalar particle $S$
with the third weak isospin charge $T^{(S)}_3$ and non-trivial electric charge $Q_S$,\footnote{We use the convention
$Q = T_{3} + Y/2$ in this paper.} the relevant
Lagrangian involving S is given
by
\begin{equation}
\label{lagrangian}
{\cal L}_S = (D_\mu S)^\dagger (D^\mu S) - m_S^2 S^\dagger S - \lambda_{HS} (H^\dagger H)(S^\dagger S)-\tilde\lambda_{HS}(H^\dagger T^a H)(S^\dagger T^a S),
\end{equation}
where $H$ is the SM Higgs doublet; $T^a$ is the $\mathrm{SU}(2)_L$ generators, and the covariant derivative involving $\gamma$ and $Z$ gauge fields is defined as
\begin{equation}
D_\mu S = (\partial_\mu + ie QA_\mu +i e g_{ZSS} Z_{\mu} )S\,,
\end{equation}
where $g_{ZSS}= (T^{(S)}_3-Q_S s_W^2)/(s_W c_W)$
with $s_W=\sin\theta_W$ and $c_W=\cos\theta_W$ ($\theta_W$ being the Weinberg mixing angle). 
We note that 
 the  mass splittings at tree level among the components of the scalar 
$S$ come from the last term in Eq.~(\ref{lagrangian}), 
while they also receive loop-induced contributions of $\cal{O}$(100)~MeV due to the gauge boson exchange diagrams~\cite{Cirelli:2005uq},
which can be ignored in our discussion. On the other hand, the 
electroweak precision measurements, that is, the oblique parameters would constrain the mass differences $\delta m$ among 
the scalar multiplet~\cite{Lavoura:1993nq} to be 
smaller than a few $\cal{O}$(10)~GeV~\cite{Baak:2012kk,Chun:2012jw}. 
As a result, we assume the masses of the scalar multiplet are degenerate in this paper for simplicity.
The trilinear 
coupling between Higgs and the scalar particle is
$-g_{hSS}h S^\dagger S =-\lambda_{HS}v h S^\dagger S$ after
Higgs develops the vacuum expectation value, $\langle H \rangle = v/\sqrt{2}$.
The amplitudes for the three Feynman diagrams shown in Fig.~\ref{FeynDiag} can be easily written down:
\begin{figure}[tbp]
\centering
\includegraphics[width=1.0\textwidth]{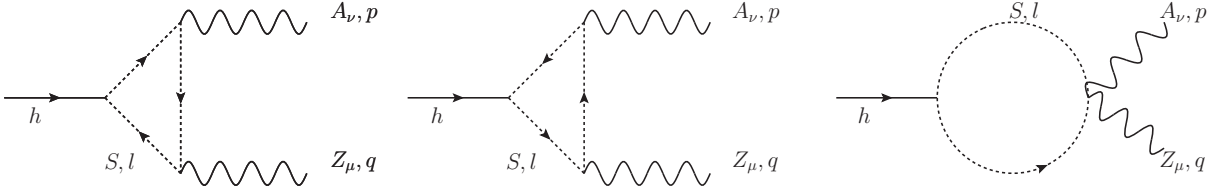}
\caption{Feynman diagrams for a charged scalar particle ($S$) contributing to $h\to Z \gamma $. \label{FeynDiag}}
\end{figure}
\begin{equation}
\begin{split}
-i {\cal M}^{(1)} &=  e^2 Q g_{ZSS}\lambda_{HS}v \mu^{4-d}\int\frac{d^d l}{(2\pi)^d}\frac{(2l+q)^{\mu}(2l+2q+p)^{\nu}\epsilon^{Z*}_\mu(q) \epsilon^{A*}_\nu(p)}{(l^2- m_S^2)[(l+q)^2-m_S^2][(l+q+p)^2-m_S^2]}\,, \\
-i{\cal M}^{(2)} &=  e^2 Q g_{ZSS}\lambda_{HS}v \mu^{4-d}\int\frac{d^d l}{(2\pi)^d}\frac{(2l-q)^{\mu}(2l-2q-p)^\nu \epsilon^{Z*}_\mu(q) \epsilon^{A*}_\nu(p)}{(l^2- m_S^2)[(l-q)^2-m_S^2][(l-q-p)^2-m_S^2]}\,, \\
-i{\cal M}^{(3)} &= -2e^2 Q g_{ZSS} \lambda_{HS} v \mu^{4-d}\int\frac{d^d l}{(2\pi)^d}\frac{g^{\mu\nu}}{(l^2-m_S^2)[(l+p+q)^2-m_S^2]} \epsilon^{Z*}_\mu(q) \epsilon^{A*}_\nu(p)\,,
\end{split}
\end{equation}
where we have taken the dimensional regularization. Note that by changing the integration variable
$l \to -l$,  $-i{\cal M}^{(2)}$ can be transformed into the form that is exactly identical to $-i{\cal M}^{(1)}$,
so that we only need to the calculate  $2(-i{\cal M}^{(1)})+(-i{\cal M}^{(3)})$. With the standard procedures of
the Feynman parametrization and the translation of the integral momentum $l$, we obtain:
\begin{equation}
\label{eq2.4}
\begin{split}
& -i(2{\cal M}^{(1)}+{\cal M}^{(3)}) \\
&= e^2 (2Q) g_{ZSS}\lambda_{HS}v \mu^{4-d}\int\frac{d^d l}{(2\pi)^d} \frac{\{(2l+q)^\mu(2l+2q+p)^\nu-[(l+q)^2-m_S^2]g^{\mu\nu}\}\epsilon^{Z*}_\mu(q) \epsilon^{A*}_\nu(p)}{(l^2- m_S^2)[(l+q)^2-m_S^2][(l+q+p)^2-m_S^2]} \\
&= e^2 (2Q) g_{ZSS}\lambda_{HS}v \mu^{4-d}\Gamma(3)\int dx dy\int\frac{d^d l}{(2\pi)^d}
\left[\frac{4l^\mu l^\nu}{(l^2-\Delta^2)^3}-g^{\mu\nu}\frac{1}{(l^2-\Delta^2)^2}\right. \\
& \left. + \frac{-4y(1-x-y)(p^\mu q^\nu-g^{\mu\nu}p\cdot q )-(1-2x-2y)(1-x-y) m_Z^2g^{\mu\nu}}{(l^2-\Delta^2)^3}\right]\epsilon^{Z*}_\mu(q) \epsilon^{A*}_\nu(p),
\end{split}
\end{equation}
where $\Delta^2\equiv m_S^2-x(1-x-y)m_Z^2-y(1-x-y)m_h^2$. In the second line of Eq.~(\ref{eq2.4}), we have used 
the following on-shell identities:
\begin{equation}
p^2=0,\quad q^2=m_Z^2,\quad (p+q)^2=m_h^2,\quad p\cdot \epsilon^{A*}(p)=q\cdot \epsilon^{Z*}(q)=0.
\end{equation}
With the help of the dimensional regularization, it can be proved that in the second equation in Eq.~(\ref{eq2.4}),
the first two terms exactly cancel with each other and
only the last finite term is left, given by
\begin{equation}
\label{eq2.6}
\begin{split}
& -i(2{\cal M}^{(1)}+{\cal M}^{(3)}) \\
&= \frac{ i e^2 (2Q) g_{ZSS}\lambda_{HS}v}{16\pi^2} \mu^{4-d}\int dx dy
\left\{\frac{4y(1-x-y)(p^\mu q^\nu-g^{\mu\nu}p\cdot q )}{m_S^2-x(1-x-y)m_Z^2-y(1-x-y)m_h^2}\right. \\
& \left. +\frac{(1-2x-2y)(1-x-y) m_Z^2g^{\mu\nu}}{m_S^2-x(1-x-y)m_Z^2-y(1-x-y)m_h^2}\right\}\epsilon^{Z*}_\mu(q) \epsilon^{A*}_\nu(p).
\end{split}
\end{equation}
 From this equation after the integration, the first term leads to
\begin{equation}
-i(2{\cal M}^{(1)}+{\cal M}^{(3)}) = \frac{ i e^2 (2Q) g_{ZSS}\lambda_{HS}v}{16\pi^2}(p^\mu q^\nu-g^{\mu\nu}p\cdot q )\epsilon^{Z*}_\mu(q) \epsilon^{A*}_\nu(p) A^{Z\gamma}_0(\tau_{S},\lambda_{S}),
\end{equation}
while the second one vanishes, where
$\tau_S=4 m_S^2/m_h^2$, $\lambda_S=4m_S^2/m_Z^2$ and the loop function $A^{Z\gamma}_0(x,y)$
is defined in Appendix~\ref{loop func}. By combining the SM contributions and integrating the phase space of outgoing particles,
we obtain the $h \rightarrow Z\gamma$ decay width,
\begin{equation}
\label{ampZgamma}
\Gamma(h\to Z\gamma) = \frac{\alpha^2}{512\pi^3}m_h^3\left(1-\frac{m_Z^2}{m_h^2}\right)^3\left|{\cal A}_{SM}^{Z\gamma}
- \frac{\lambda_{HS}v}{m_S^2}(2 \sum_{T_3}Q_S\cdot g_{ZSS})A_0^{Z\gamma}(\tau_S,\lambda_S)\right|^2,
\end{equation}
with
\begin{equation}
\quad\quad\quad {\cal A}_{SM}^{Z\gamma}= \frac{2}{v}\left[\cot\theta_W A_1^{Z\gamma}(\tau_W,\lambda_W)
+ N_c\frac{(2Q_t)(T_3^{(t)}-2Q_t s_W^2)}{s_W c_W}A_{1/2}^{Z\gamma}(\tau_t,\lambda_t)\right]\,,
\end{equation}
where $\tau_i=4m_i^2/m_h^2$, $\lambda_i=4m_i^2/m_Z^2$ ($i=W,t$), and
the summation is over the different isospin components in  given $SU(2)_L$ multiplets. 
We point out that for the non-SM contribution to the decay rate of $h\to Z\gamma$,  
there are extra factors of -1/2  in Refs.~\cite{Carena:2012xa,Picek:2012ei} and -1 in Ref.~\cite{Huang:2012rh}
in comparing with ours in Eq.~(\ref{ampZgamma}).
The different signs for the scalar contributions to $h\rightarrow Z\gamma$  clearly
lead to different results by taking limits to $h\rightarrow \gamma\gamma$.
The modification in the partial decay width of $h\to Z\gamma$ is then expressed in terms of
the enhancement factor, given by
\begin{equation}\label{Zg}
R_{Z\gamma} \equiv {\Gamma( h\to Z\gamma)\over \Gamma_{SM}(h\to Z\gamma)}
= \left|1- \tilde{N}\lambda_{HS}\frac{v^2}{m_S^2}\left(2 \sum_{T_3}Q_S\cdot g_{ZSS}\right)\frac{A_0^{Z\gamma}(\tau_S,\lambda_S)}{v{\cal A}_{SM}^{Z\gamma}}\right|^2,
\end{equation}
where $\Gamma_{SM}(h\to Z\gamma)$ represents the decay width in the SM and
the factor $\tilde{N}$ represents the degeneracy of the multiplet.
For completeness and convenience for the later discussions,
we also present the standard formula for the enhancement factor of the diphoton rate,
\begin{equation}
\label{gg}
R_{\gamma\gamma}\equiv {\Gamma( h\to \gamma\gamma)\over \Gamma_{SM}(h\to \gamma\gamma)} = \left|1+\tilde{N}\frac{\lambda_{HS}}{2}\frac{v^2}{m_S^2}
\left(\sum_{T_3}Q_S^2\right)\frac{A^{\gamma\gamma}_0(\tau_S)}{A^{\gamma\gamma}_1(\tau_W)+N_c Q_t^2 A^{\gamma\gamma}_{1/2}(\tau_t)}\right|^2,
\end{equation}
where the loop functions $A^{\gamma\gamma}_j(x)\; (j=0,1/2,1)$ are defined in Appendix~\ref{loop func}. 
It is straightforward to show that our formula in Eq.~(\ref{Zg}) for the $Z\gamma$ decay  can be retrieved 
to the one in Eq.~(\ref{gg}) for the $\gamma\gamma$ mode when taking $m_{Z} \rightarrow 0$ and making 
the replacement $g_{ZSS}\to Q_S$. 
If the mass splittings among components of the 
multiplet are taken into account, the deviations of the contributions in Eqs.~(\ref{Zg}) and (\ref{gg}) 
are approximately proportional to $\delta m^2/m_S^2$. The modifications from the mass splittings are at percentage 
level and are negligible due to the electroweak precision measurements.
Consequently, we will simply take Eqs.~(\ref{Zg}) and (\ref{gg}) 
to analyze the qualitative behaviors in  $h\to \gamma\gamma$ and $h\to Z\gamma$ decays.

\section{Phenomenological Analyses of Correlations Between the $\gamma\gamma$ and $Z\gamma$ Channels}
\label{sec:3}
With formulae in Eqs.~(\ref{Zg}) and (\ref{gg}) at hand, we are able to study the correlations between
$h \rightarrow \gamma\gamma$ and $h \rightarrow Z\gamma$ decay rates. According to the EW gauge
symmetry, one expects that the particle responsible for the observed enhancement in the diphoton channel could
also lead to the modification of the $Z\gamma$ rate.
Furthermore, it is interesting to notice that the $Z\gamma$ amplitude depends not only on the electric charges of
particles running in the loops but also on their isospins, in contrast with the $\gamma\gamma$ case in which only
electric charges can be probed. Consequently, a combined analysis of the decay widths in these two modes can provide
us with important information on the EW charges of particles contributing to both processes.

\subsection{Singlet and Doublet Scalars}
\begin{figure}[tbp]
\centering
\includegraphics[width=75mm,height=60mm]{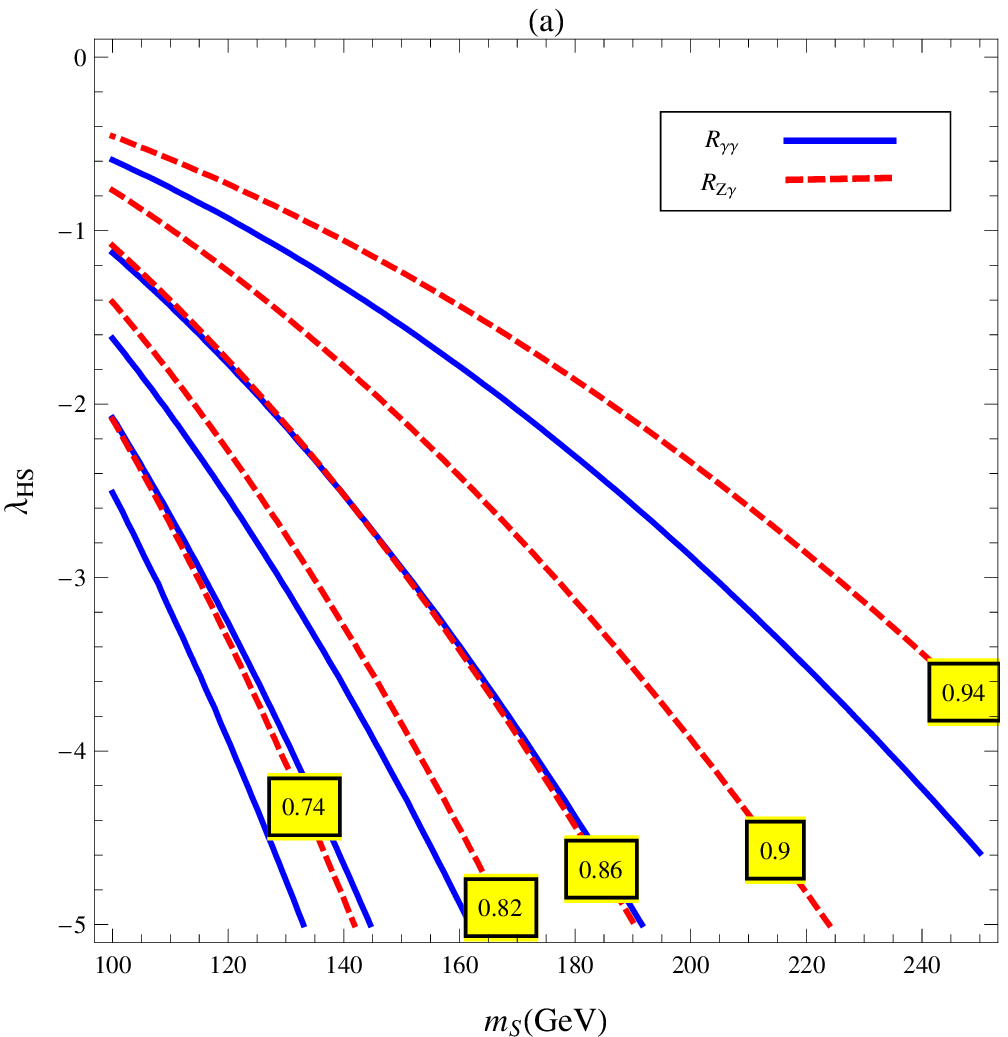}
\includegraphics[width=75mm,height=60mm]{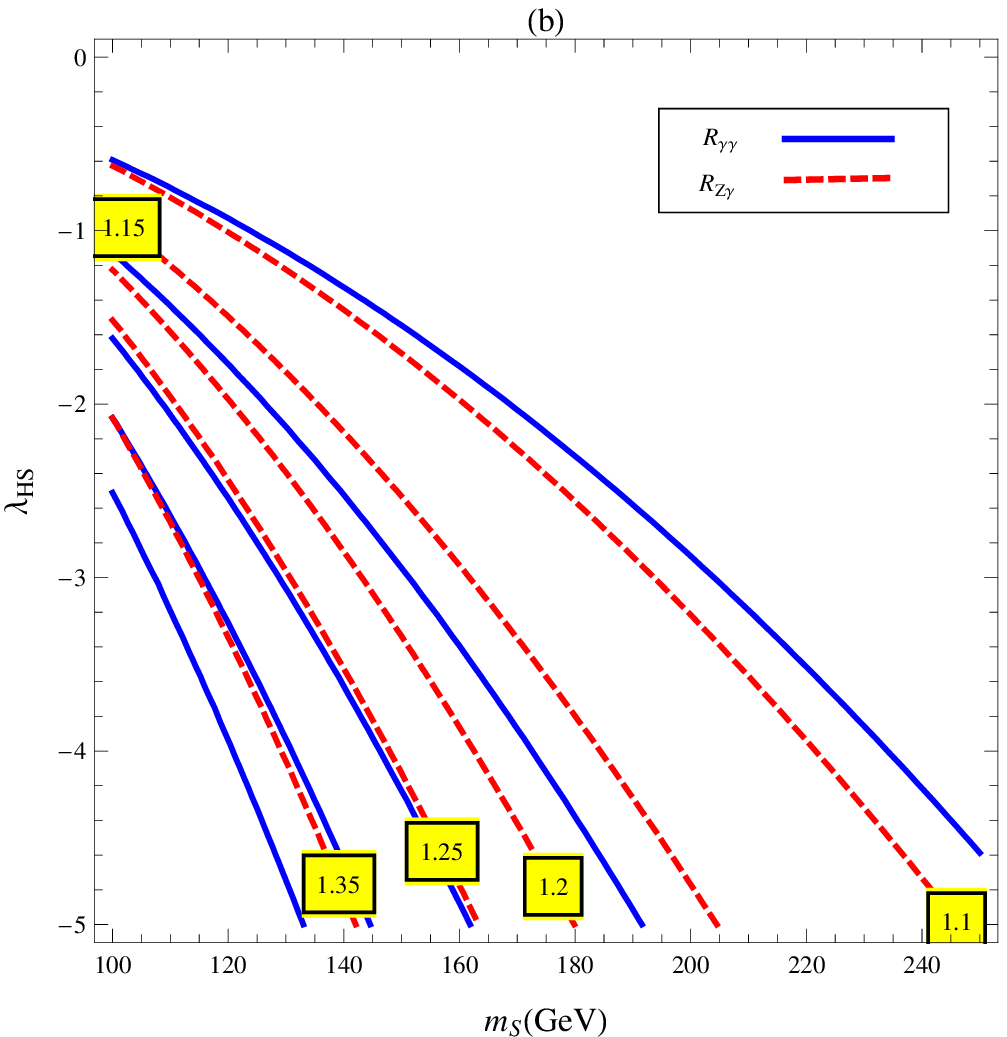}
\caption{Contours of $R_{\gamma\gamma}$ (solid lines) and $R_{Z\gamma}$ (dashed lines with the values in the yellow boxes)
 in $m_S$-$\lambda_{HS}$ plane
for scalars of (a) singlet (T=0 and Y=2)
and (b) doublet (T=1/2 and Y=1), where
 the solid (dashed) lines from top to
bottom correspond to the values of (a) 1.25 (0.94), 1.50 (0.9), 1.75 (0.86), 2.00 (0.82) and 2.25 (0.74); and (b)
1.25 (1.10), 1.50 (1.15), 1.75 (1.20), 2.00 (1.25) and 2.25 (1.35), respectively.
}\label{ZGL}
\end{figure}

In this subsection, we revisit  two simple cases of singlet and doublet scalars under  $SU(2)_L$
with the  couplings:
\begin{equation}
\label{2cases}
({\rm I}):\quad g^{(1)}_{ZSS} = \frac{-Q_S s_W^2}{s_W c_W},\quad  \quad ({\rm II}): \quad g^{(2)}_{ZSS}=\frac{1}{s_W c_W} \left(\frac{1}{2}-Q_S s_W^2\right),
\end{equation}
where $Q_S=1$, corresponding to $(T,Y)=(0,2)$ and $(1/2,1)$, respectively. The aim for this part of the 
investigation is twofold: (a) illustrating the features of the isospin and hypercharge dominations and (b) reexamining 
the results in Ref.~\cite{Carena:2012xa}.
In the two cases in Eq.~(\ref{2cases}), since only charged particles with $Q_{S} = 1$ run in the loops, the 
enhancements for the diphoton rate are expected to be  the same. However, due to  the different isospin 
representations, those for the $Z\gamma$ mode are distinct.  From  Fig.~\ref{ZGL}, we see that the 
$Z\gamma$ rate in Case I (II) has a small suppression (enhancement) in comparison with the SM value.
These features can be understood from the general formula in Eq.~(\ref{Zg}).
 In order to enhance the $h\to\gamma\gamma$ rate by a constructive
interference, it is required that  $\lambda_{HS}$ must be negative since $A_0^{\gamma\gamma}(\tau_S)$
and $A^{\gamma\gamma}_{SM}$ possess an opposite sign.
In this case, the $h\to Z\gamma$ rate
depends only on the sign of $Q_S g_{ZSS}$ as $\lambda_{HS}<0$  corresponds to both positive values of ${\cal A}^{Z\gamma}_{SM}$
and $A^{Z\gamma}_0(\tau_S,\lambda_S)$ in the physically interesting mass regime.
As a result, $R_{Z\gamma}$ is enhanced (suppressed) if
  $Q_S g_{ZSS}$ is positive (negative).
 Accordingly, it is easy to see that $Q_S g_{ZSS} < 0\, (>0) $ in Case I (II).
Note that our result is opposite to that in Ref.~\cite{Carena:2012xa}, 
in which the scalar contributions to $h\rightarrow\gamma\gamma$ and $h\rightarrow Z\gamma$ 
are positive correlated in Case I and negative correlated in Case II due to the extra minus sign in the formula.

The factor $Q_S \cdot g_{ZSS}$ can be examined in a general  scalar multiplet.
By using the identity $Q=T_3+Y/2$, we have
\begin{equation}
Q_S \cdot g_{ZSS} = \frac{1}{s_W c_W}\left(T_3+\frac{Y}{2}\right)\left(T_3 c_W^2-\frac{Y s_W^2}{2}\right).
\end{equation}
When the isoweak charge $T$ of the scalar multiplet is larger than its hypercharge $Y$, we find $Q_S\cdot g_{ZSS}>0$, 
which indicates the enhanced behavior in the $Z\gamma$ channel compared with the SM prediction. In contrast, for the 
multiplet with $Y \gg T$, we get $Q_S\cdot g_{ZSS}<0$, leading to a suppressed  rate of  $h\to Z\gamma$. 
Precisely, Cases I and II are the two simple but nontrivial archetypes of the two limits, respectively.

\subsection{Triplet Scalars with $T=1$ and $Y=0, 2$}
\begin{figure}[t]
\centering
\includegraphics[width=75mm,height=60mm]{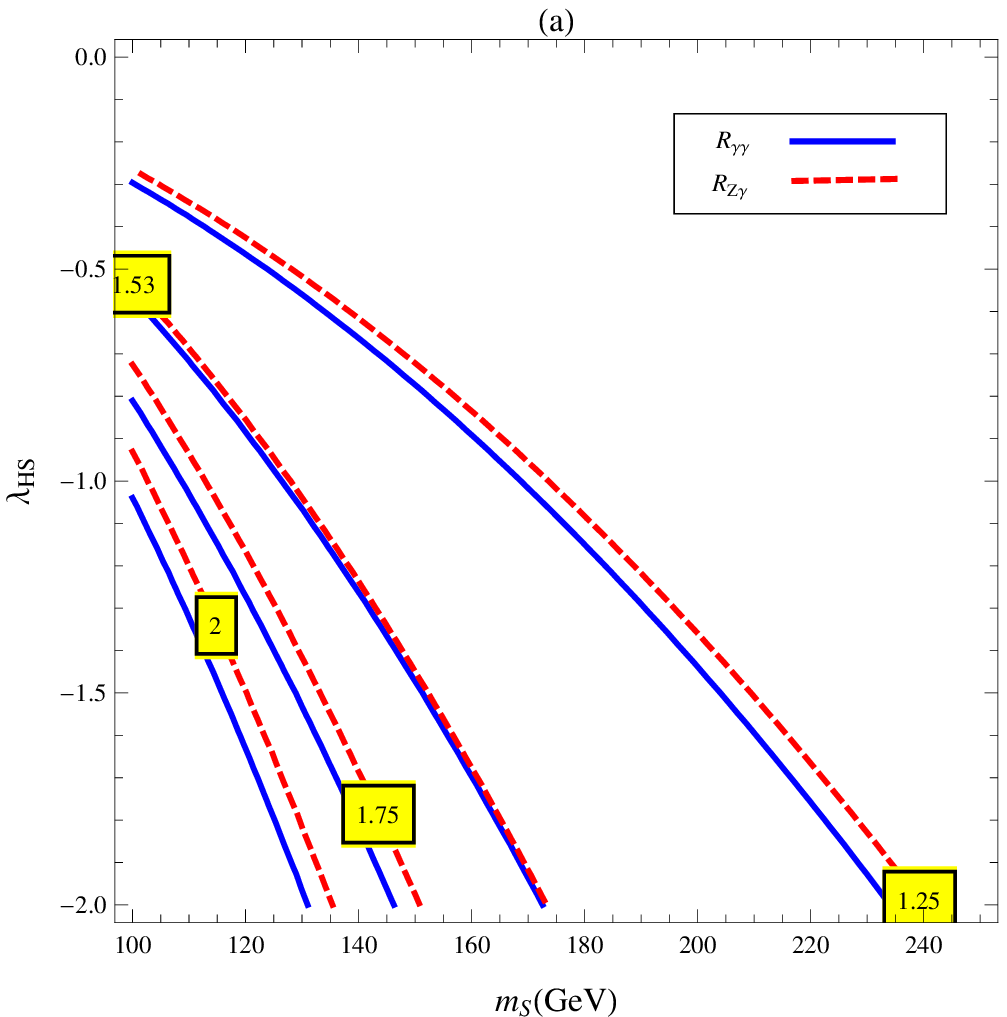}
\includegraphics[width=75mm,height=60mm]{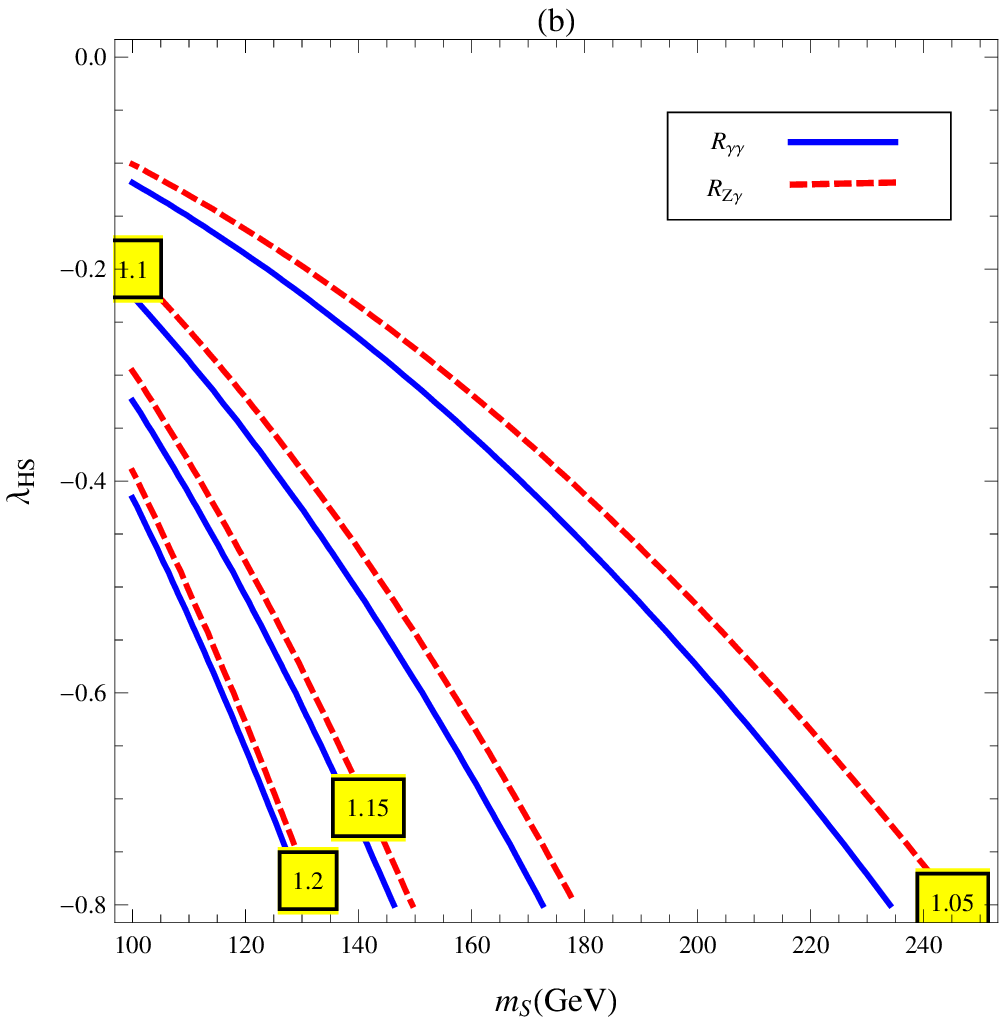}
\caption{Contours of $R_{\gamma\gamma}$ (solid lines) and $R_{Z\gamma}$ (dashed lines) in $m_S$-$\lambda_{HS}$ plane
for the triplet scalars of (a) (T,Y)= (1,0)
and (b) (T,Y)=(1,2), where
 the solid (dashed) lines from top to
bottom correspond to the values of (a) 1.25 (1.25), 1.50 (1.53), 1.75 (1.75), and 2.00 (2.00); and (b)
1.25 (1.05), 1.50 (1.10), 1.75 (1.15), and 2.00 (1.20), respectively.
}\label{ZG3}
\end{figure}

It is known that triplet scalars are introduced in many SM extensions,
such as the type II seesaw model~\cite{Magg:1980ut,Schechter:1980gr,Cheng:1980qt,Gelmini:1980re,Lazarides:1980nt,Mohapatra:1980yp,Schechter:1981cv}.
Current LHC experiments also plan to search for these triplet scalars, especially the doubly charged particle in the
$Y=2$ case due to its interesting signatures~\cite{Chatrchyan:2012ya,ATLAS:2012hi}.
Instead of a special model, we examine the generic feature of these triplets. In particular, we
investigate their loop effects in the $\gamma\gamma$ and $Z\gamma$ decay channels of the Higgs particle.
For simplicity,
we assume the multiplets of scalars possess the same masses and no mixings between the triplet scalars and the SM Higgs.
For more detail analysis one can refer to Refs.~\cite{Melfo:2011nx,Arhrib:2011vc,Akeroyd:2012ms,Chun:2012jw,Dev:2013ff}. Our result is shown in Fig.~\ref{ZG3}.
 Clearly, the $Z\gamma$ rate is enhanced in both cases, which belong to the isospin domination class
discussed in the previous subsection. Furthermore, the enhancement in the $Y=0$ case is much larger than that in the $Y=2$
one, which implies  that increasing the value of $Y$ will decrease the $h\to Z\gamma$ decay rate for a given isospin multiplet.
Moreover, in comparison  with Case I of $(T,Y)=(0,2)$
in Sec. III-A,  $\Gamma(h\to Z\gamma)$ for the case of  $(T,Y)=(1,2)$
 for a given rate of $h\to \gamma\gamma$, say, 1.5 times the SM value, is slightly larger, showing that
a large  $SU(2)_{L}$ multiplet helps to increase the $h\to Z\gamma$ rate.

\subsection{5-plet Scalars with $T=2$ and $Y=0,2$}
\begin{figure}[t]
\centering
\includegraphics[width=75mm,height=60mm]{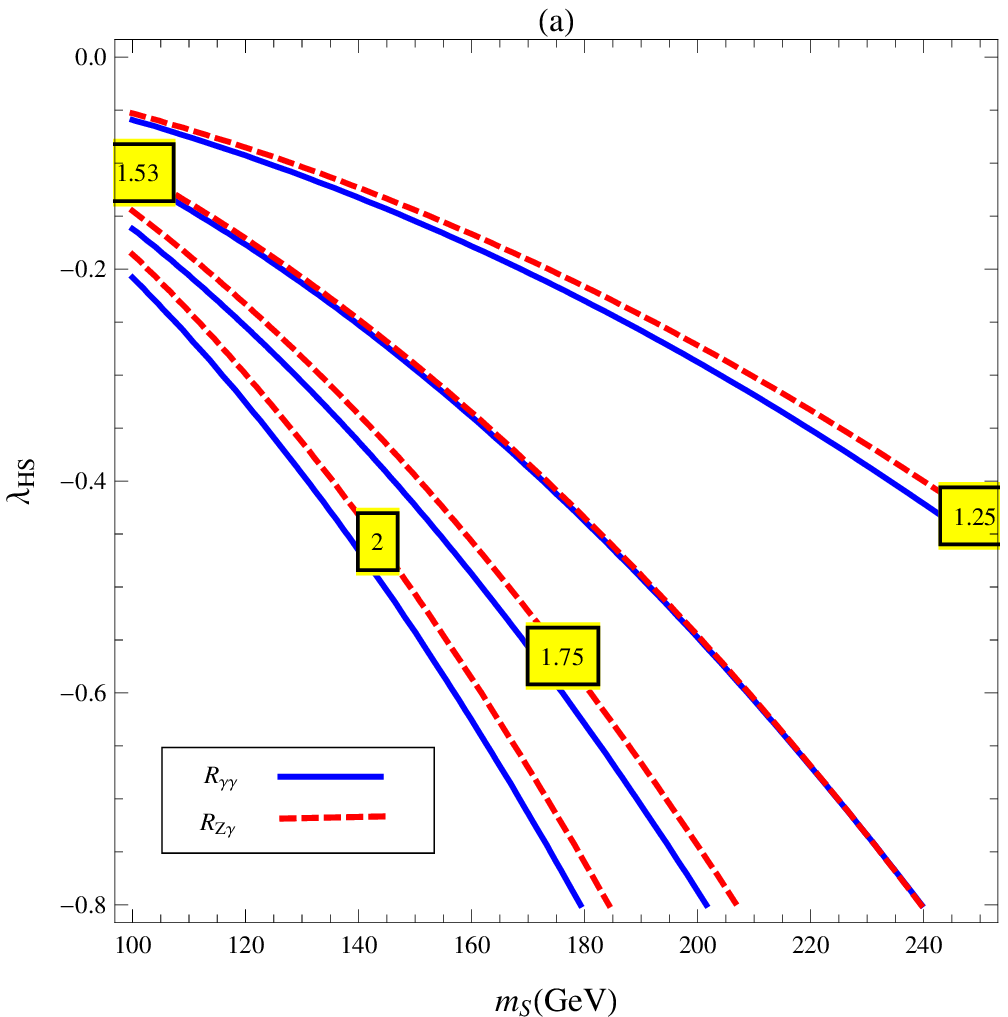}
\includegraphics[width=75mm,height=60mm]{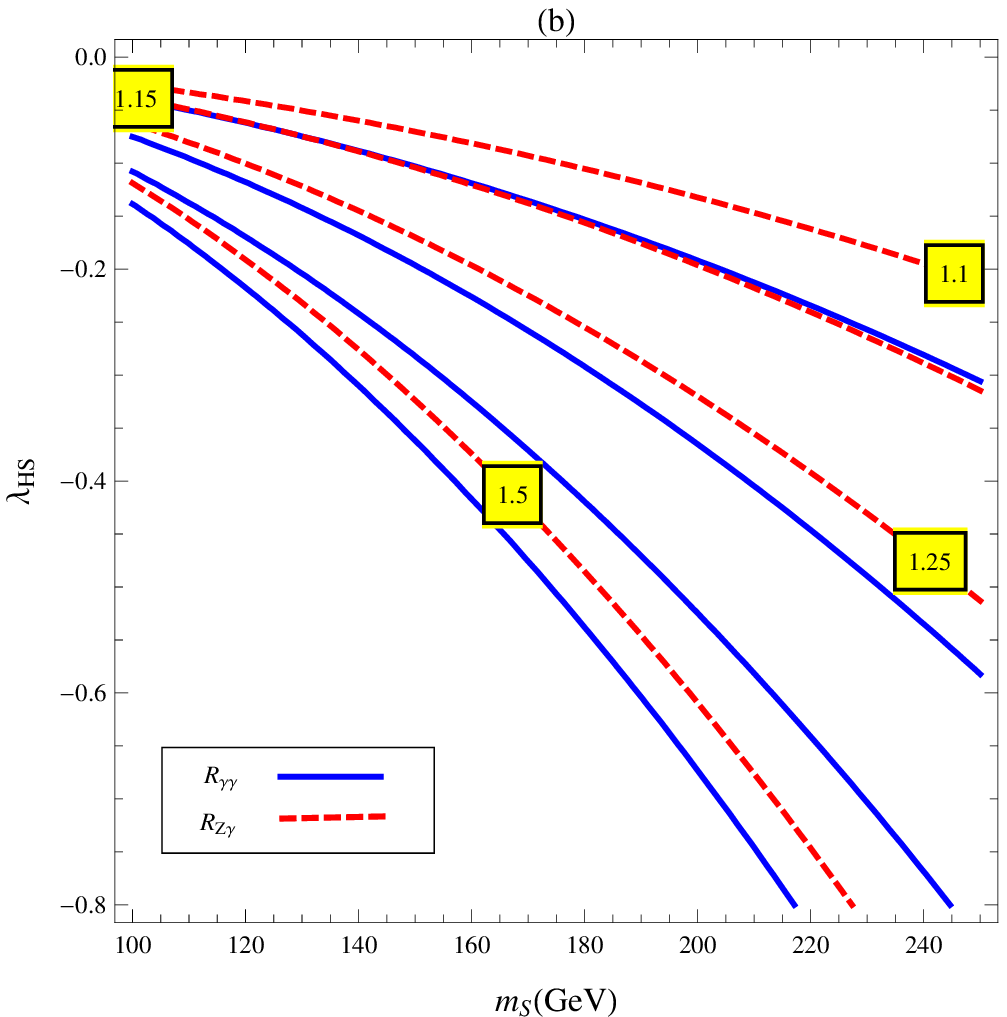}
\caption{Contours of $R_{\gamma\gamma}$ (solid lines) and $R_{Z\gamma}$ (dashed lines) in $m_S$-$\lambda_{HS}$ plane
for the 5-plet scalars of (a) (T,Y)= (2,0)
and (b) (T,Y)=(2,2), where
 the solid (dashed) lines from top to
bottom correspond to the values of (a) 1.25 (1.25), 1.50 (1.53), 1.75 (1.75), and 2.00 (2.00); and (b)
1.25 (1.10), 1.50 (1.15), 1.75 (1.25), and 2.00 (1.50), respectively.
}\label{ZG5}
\end{figure}

The weak $SU(2)_{L}$ 5-plet scalar particles are also of great phenomenological interest. For example, a recent study
in Ref.~\cite{Chen:2012vm} shows that with the 5-plet and $Y=2$ scalar, the small Majorana neutrino masses can be generated at two-loop
level. The $\gamma\gamma$ and $Z\gamma$ results are presented in Fig.~\ref{ZG5}.
Similar to the $SU(2)_{L}$ triplet cases, the 5-plet scalars give sizable enhancements in the $h\to Z\gamma$ channel
when the diphoton rate is lying in 1.5 to 2.0 times larger than the SM value.
Explicitly,
the enhancement factor $R_{Z\gamma}$  is also about 1.5 to 2.0 (1.25 to 1.50) for the $Y=0\, (2)$ case.
Again, the behaviors of these two cases can be expected by the rules of the isospin
domination. We note that a bit less enhancement in the $Y=2$ case is due to the hypercharge cancellation.

\subsection{General Discussions}
\begin{figure}[tbp]
\centering
\includegraphics[scale=0.8]{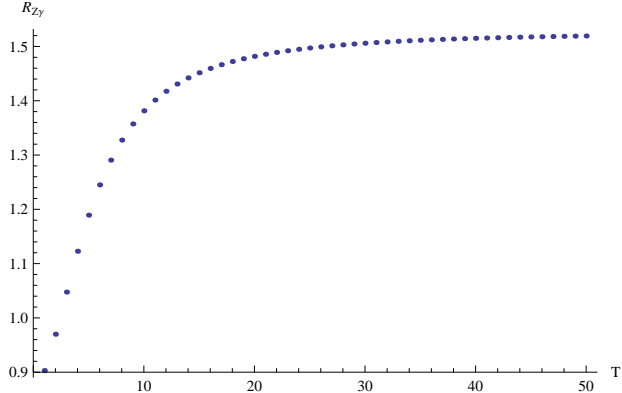}
\caption{ $R_{Z\gamma}$ as a function of the isospin $T$ with
$R_{\gamma\gamma}=1.5$ and  $m_S=300$~GeV.}\label{changeT}
\end{figure}

In this subsection, we give a more comprehensive overview of how to use the combined analysis of
the $\gamma\gamma$ and $Z\gamma$ modes to investigate the nature of the new scalar multiplets.
By assuming that the excess of $h\to \gamma\gamma$  is originated from a constructive interference between the SM and charged scalar loops,
we  obtain
\begin{equation}\label{general}
\begin{split}
\frac{\sqrt{R_{Z\gamma}}-1}{\sqrt{R_{\gamma\gamma}}-1} &= -\frac{4 [A^{\gamma\gamma}_1(\tau_W)
+ N_c Q_t^2 A^{\gamma\gamma}_{1/2}(\tau_t)]}{v{\cal A}^{Z\gamma}_{SM}}
\frac{A_0^{Z\gamma}(\tau_S, \lambda_S)}{A_0^{\gamma\gamma}(\tau_S)}
\frac{\sum_{T_3} Q_S\cdot g_{ZSS}}{\sum_{T_3}Q^2_S} \\
&= 2.71\cdot \frac{A_0^{Z\gamma}(\tau_S, \lambda_S)}{A_0^{\gamma\gamma}(\tau_S)}
\frac{4T(T+1)c_W^2-3Y^2 s_W^2}{4T(T+1)+3Y^2}
\end{split}
\end{equation}
for a scalar with general charges of $T$ and $Y$.
 From Eq.~(\ref{general}),
   $R_{Z\gamma}$ increases with a higher value\footnote{The renormalization group running will
drive the $SU(2)_{L}$ gauge coupling to violate perturbativity for $2T + 1 > 8$~\cite{Cirelli:2005uq}. The
main point in this paper is to show the quantitative behaviors of the scalars with arbitrary EW quantum numbers.} of $T$
since $A_0^{Z\gamma}$ and $A_0^{\gamma\gamma}$ are both positive, whereas
 it decreases with a larger $|Y|$.
Furthermore, when $T\gg |Y|$,
$R_{Z\gamma}$ is saturated to a maximal value with a fixed value of $R_{\gamma\gamma}$, given by
\begin{equation}\label{LargeT}
\frac{\sqrt{R^{max}_{Z\gamma}}-1}{\sqrt{R_{\gamma\gamma}}-1}
=2.09\cdot
\frac{A_0^{Z\gamma}(\tau_S, \lambda_S)}{A_0^{\gamma\gamma}(\tau_S)},
\end{equation}
which is definitely positive, implying that the $Z\gamma$ decay width must be enhanced. This limit corresponds
to the isospin domination case as discussed in Sec. III-A.
In Fig.~\ref{changeT}, we
show $R_{Z\gamma}$ as a function of the isospin $T$ with $R_{\gamma\gamma}=1.5$ and
the scalar mass $m_S=300$~GeV. From the figure, we find that the saturated value of $R^{max}_{Z\gamma}$ is $1.53$.

\begin{figure}[tbp]
\centering
\includegraphics[scale=0.8]{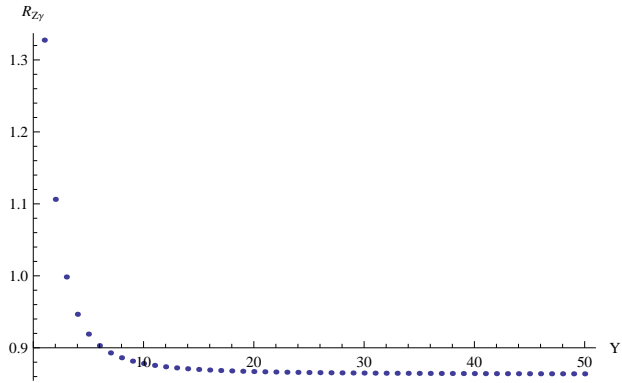}
\caption{$R_{Z\gamma}$ as a function of the hypercharge $Y$
with $R_{\gamma\gamma}=1.5$ and  $m_S=300$~GeV.}\label{changeY}
\end{figure}

On the other hand, when we take the large hypercharge limit $|Y|\gg T$, that is, the hypercharge domination case, we
obtain the minimum value for the factor $R_{Z\gamma}$.
\begin{equation}\label{LargeY}
\frac{\sqrt{R^{min}_{Z\gamma}}-1}{\sqrt{R_{\gamma\gamma}}-1}
= -0.63\cdot\frac{A_0^{Z\gamma}(\tau_S, \lambda_S)}{A_0^{\gamma\gamma}(\tau_S)}.
\end{equation}
In Fig.~\ref{changeY}, we show  $R_{Z\gamma}$ as a function of the hapercharge $Y$
with $R_{\gamma\gamma}=1.5$ and $m_S=300$~GeV.
In the large-$|Y|$ limit, we get  the asymptotic value of $R^{min}_{Z\gamma}=0.86$.

\begin{figure}[tbp]
\centering
\includegraphics[width=80mm,height=65mm]{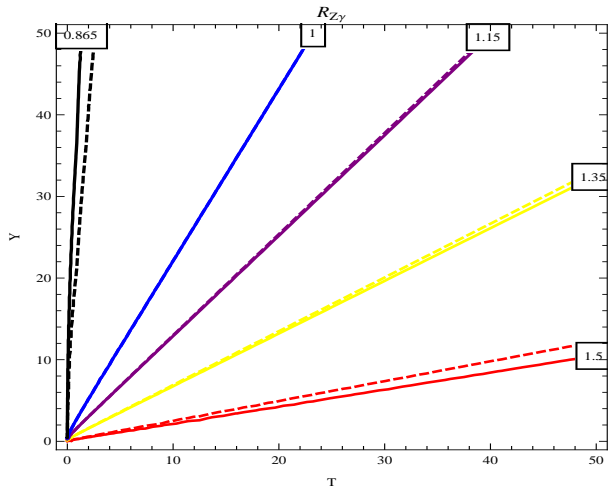}
\caption{$R_{Z\gamma}$ in the plane of general $SU(2)_{L}\times U(1)_{Y}$ quantum numbers
by fixing $R_{\gamma\gamma}=1.5$, where the solid and dashed lines represent the results of  $m_S\to \infty$ and
 $m_S=200$~GeV, respectively.}\label{RLM}
\end{figure}

The above features become more transparent in the large scalar mass limit, $m_S^2\to \infty$.
In this limit, we have that $A_0^{Z\gamma}(\tau_S, \lambda_S)=A_0^{\gamma\gamma}(\tau_S)/2=1/6$,
and thus Eq.~(\ref{general}) can be reduced to
\begin{equation}\label{LargeMass}
\frac{\sqrt{R_{Z\gamma}}-1}{\sqrt{R_{\gamma\gamma}}-1}= 1.36\cdot
\frac{4T(T+1)c_W^2-3Y^2 s_W^2}{4T(T+1)+3Y^2}.
\end{equation}
In most of the physically interesting regime, Eq.~(\ref{LargeMass}) is a very good approximation with the
error of ten percent for the scalar mass above 200~GeV in which the ratio
$\frac{A_0^{Z\gamma}(\tau_S,\lambda_S)}{A_0^{\gamma\gamma}(\tau_S)}$ is numerically stable around $0.5$.
A remarkable feature shown in Eq.~(\ref{LargeMass}) is that the $h\to Z\gamma$ decay rate for the Higgs boson only depends
on the EW quantum numbers assigned to the scalar multiplet for a given enhancement in the $\gamma\gamma$ channel.
As an illustration, Fig.~\ref{RLM} shows  the contours of $R_{Z\gamma}$  in terms of  the general $SU(2)_{L}\times U(1)_{Y}$
quantum numbers with $R_{\gamma\gamma}=1.5$ for both $m_{S} \rightarrow \infty$
and $m_S = 200$~GeV, respectively.
In the large mass regime, the maximum and minimum values of $R_{Z\gamma}$ can be easily obtained,
$R_{Z\gamma}^{max}= [1+1.04(\sqrt{R_{\gamma\gamma}}-1)]^2$ and $R_{Z\gamma}^{min} =[1-0.31(\sqrt{R_{\gamma\gamma}}-1)]^2$,
corresponding to the isospin and hypercharge dominations, respectively.

Finally, we can further demonstrate that the above bounds hold when we extend our discussion to arbitrary numbers
of scalar multiplets with different quantum numbers, masses $m_i$, Higgs couplings $\lambda_i$, and
the degeneracies $N_i$. The formula in Eq.~(\ref{general}) can be generalized to:
\begin{equation}\label{MoreGeneral}
\frac{\sqrt{R_{Z\gamma}}-1}{\sqrt{R_{\gamma\gamma}}-1} = 1.36\cdot
\frac{\sum_i N_i\frac{|\lambda_i|}{m_i} (2T_i+1)[4T_i(T_i+1)c_W^2 -
3Y_i^2 s_W^2]}{\sum_i N_i\frac{|\lambda_i|}{m_i}(2T_i+1)[4T_i(T_i+1)+3Y_i^2]}.
\end{equation}
where we have taken $m_{i} > 200$~GeV.
Note that Eq.~(\ref{MoreGeneral}) is valid only for
 the new scalar contributions interfering with the SM part constructively. This assumption follows by two indications:
(i)  $\lambda_i$ should be negative in sign, so that we can replace $\lambda_i$ with its absolute
value in Eq.~(\ref{MoreGeneral}); and (ii)
contributions to the amplitudes of the $h\to\gamma\gamma$ and
$Z\gamma$ modes must be smaller than the corresponding SM parts in size to eliminate the sign ambiguity when we take the square
roots of  $R_{\gamma\gamma}$ and $R_{Z\gamma}$.
We now try to derive a generic bound for $R_{Z\gamma}$.
 From Eq.~(\ref{MoreGeneral}) and inequalities
\begin{equation}
 -0.31\left[4T_i(T_i+1)+3Y_i^2\right] < 4T_i(T_i+1)c_W^2-3Y_i^2 s_W^2
 < 1.04 \left[4T_i(T_i+1)+3Y_i^2\right]~~\;
\end{equation}
 for a single multiplet,
we find
\begin{equation}
\label{GB}
-0.31<\frac{\sqrt{R_{Z\gamma}}-1}{\sqrt{R_{\gamma\gamma}}-1}<1.04.
\end{equation}
From the general bound in Eq.~(\ref{GB}), it is clear that one can deduce the allowed range of the $Z\gamma$ 
decay width for a given value of $R_{\gamma\gamma}$. Since the current ATLAS and CMS data have indicated 
the range $1.5<R_{\gamma\gamma}<2.0$ for the diphoton mode, one predicts that
\begin{equation}
\label{ZGB}
0.76<R_{Z\gamma}<2.05
\end{equation}
for the $Z\gamma$ mode, where both the upper and lower limits arise
from $R_{\gamma\gamma}=2.0$. We emphasize  that the result in  Eq.~(\ref{ZGB}) does not rely on the details of the
extra scalars, such as the quantum numbers and their couplings to the Higgs particle.
Thus, if the future-measured diphoton rate is still enhanced about $1.5-2.0$ times over the SM value,
the $Z\gamma$ decay rate should be in the range of 0.76 to 2.05. Otherwise, our simple scenario discussed in this paper will be excluded.

\section{Summary}
\label{sec:4}
We have studied the decay width of  the $h\to Z\gamma$ mode with extra scalar multiplets of arbitrary weak isospins
and hypercharges. Due to the SM gauge symmetry, these new scalars which result in the excess of the $h\to \gamma\gamma$ decay rate would
generically shift the  $h\to Z\gamma$ decay rate as well.
The combined analysis of these two modes can provide us with valuable information about the new physics structure.
Regarding the ambiguity of the  formulae for $\Gamma(h\to Z\gamma)$ in the literature,
we have revisited some simple extensions to the SM in which new scalar multiplets are introduced.
We have found that $\Gamma(h\to Z\gamma)$ only depends on the relative size of the isospin $T$ and the absolute value of
the hypercharge $Y$. In particular, we have shown that  the enhancement factor  $R_{Z\gamma}$ is a monotonically increasing function of $T$
and a monotonically decreasing one of $|Y|$ with the fixed value of $R_{\gamma\gamma}$. This observation
enables us to predict that $0.76<R_{Z\gamma}<2.05$ by imposing
the observed range of $1.5<R_{\gamma\gamma}<2$ if the scalars are heavier
than 200~GeV. Note that this range is irrelevant to the number of the scalar multiplets, their representations, and the
couplings to the Higgs particle.
Our results on the $h\to Z\gamma$ decay clearly can be tested at the LHC.

\appendix
\section{Definition of Loop Functions}\label{loop func}
The functions related to $h \rightarrow \gamma\gamma$ and $h \rightarrow Z\gamma$ are defined as follows:
\begin{subequations}
\begin{align}
A_1^{\gamma\gamma}(x) &= -x^2[2x^{-2}+3x^{-1}+3(2x^{-1}-1)f(x^{-1})],\\
A_{1/2}^{\gamma\gamma} (x) &= 2x^2[x^{-1}+(x^{-1}-1)f(x^{-1})],\\
A_0^{\gamma\gamma} (x) &= -x^2[x^{-1}-f(x^{-1})],\\
A_1^{Z\gamma}(x,y) &= 4(3-\tan^2\theta_W)I_2(x,y)+[(1+2x^{-1})\tan^2 \theta_W-(5+2x^{-1})]I_1(x,y),\\
A_{1/2}^{Z\gamma} (x,y) &= I_1(x,y)-I_2(x,y),\\
A_0^{Z\gamma}(x,y) &= I_1(x,y),
\end{align}
\end{subequations}
where
\begin{subequations}
\begin{align}
I_1(x,y) &= \frac{x y}{2(x-y)}+\frac{x^2 y^2}{2(x-y)^2}[f(x^{-1})-f(y^{-1})]+\frac{x^2 y}{(x-y)^2}[g(x^{-1})-g(y^{-1})],\\
I_2(x,y) &= -\frac{x y}{2(x-y)}[f(x^{-1})-f(y^{-1})].
\end{align}
\end{subequations}
For a Higgs mass smaller than twice of that of the loop particle, $i.e.$ $m_h<2 m_{\mbox{loop}}$, we have
\begin{subequations}
\begin{align}
f(x) &= \arcsin^2\sqrt{x},\\
g(x) &= \sqrt{x^{-1}-1}\arcsin\sqrt{x}.
\end{align}
\end{subequations}

\acknowledgments
The work was supported in part by National Center for Theoretical Science, National Science
Council (NSC-98-2112-M-007-008-MY3 and NSC-101-2112-M-007-006-MY3) and National Tsing-Hua
University (102N1087E1), Taiwan, R.O.C.

\end{document}